\documentclass[%
reprint,
superscriptaddress,
amsmath,
amssymb, 
aps,
prl,
nofootinbib,
]{revtex4-2}
\usepackage{graphicx}
\usepackage{dcolumn}
\usepackage{bm}
\usepackage{xcolor}
\usepackage{float}
\usepackage{multirow}
\usepackage{threeparttable}
\usepackage[colorlinks=true]{hyperref}
\usepackage[T1]{fontenc}

\newcommand\isotope[2]{\textsuperscript{#2}#1}

\usepackage[symbol]{footmisc}

\begin{document}

\preprint{APS/123-QED}

\title{Precision mass measurements of \isotope{Sr}{74-76} using TITAN's Multiple Reflection Time-of-Flight Mass Spectrometer}
\author{Z. Hockenbery}
\affiliation{TRIUMF, 4004 Wesbrook Mall, Vancouver, British Columbia V6T 2A3, Canada}
\affiliation{Department of Physics, McGill University, 3600 Rue University, Montréal, QC H3A 2T8, Canada}

\author{T. Murb\"ock}
\affiliation{TRIUMF, 4004 Wesbrook Mall, Vancouver, British Columbia V6T 2A3, Canada}
\affiliation{II.~Physikalisches Institut, Justus-Liebig-Universit{\"a}t Gie{\ss}en, 35392 Gie{\ss}en, Germany}

\author{B. Ashrafkhani}
\affiliation{TRIUMF, 4004 Wesbrook Mall, Vancouver, British Columbia V6T 2A3, Canada}
\affiliation{Department of Physics and Astronomy, University of Calgary, Calgary, Alberta T2N 1N4, Canada}

\author{J. Bergmann}
\affiliation{II.~Physikalisches Institut, Justus-Liebig-Universit{\"a}t Gie{\ss}en, 35392 Gie{\ss}en, Germany}

\author{C. Brown}
\affiliation{TRIUMF, 4004 Wesbrook Mall, Vancouver, British Columbia V6T 2A3, Canada}
\affiliation{School of Physics and Astronomy, University of Edinburgh, Edinburgh EH9 3FD, United Kingdom}

\author{T. Brunner}
\affiliation{Department of Physics, McGill University, 3600 Rue University, Montréal, QC H3A 2T8, Canada}

\author{J. Cardona}
\affiliation{TRIUMF, 4004 Wesbrook Mall, Vancouver, British Columbia V6T 2A3, Canada}
\affiliation{Department of Physics and Astronomy, University of Manitoba, Winnipeg, Manitoba R3T 2N2, Canada}

\author{T. Dickel}
\affiliation{II.~Physikalisches Institut, Justus-Liebig-Universit{\"a}t Gie{\ss}en, 35392 Gie{\ss}en, Germany}
\affiliation{GSI Helmholtzzentrum f{\"u}r Schwerionenforschung GmbH, Planckstra{\ss}e 1, 64291 Darmstadt, Germany}

\author{E. Dunling}
\affiliation{TRIUMF, 4004 Wesbrook Mall, Vancouver, British Columbia V6T 2A3, Canada}

\author{J. D. Holt}
\affiliation{TRIUMF, 4004 Wesbrook Mall, Vancouver, British Columbia V6T 2A3, Canada}
\affiliation{Department of Physics, McGill University, 3600 Rue University, Montréal, QC H3A 2T8, Canada}

\author{C. Hornung}
\affiliation{GSI Helmholtzzentrum f{\"u}r Schwerionenforschung GmbH, Planckstra{\ss}e 1, 64291 Darmstadt, Germany}

\author{B. S. Hu}
\affiliation{National Center for Computational Sciences, Oak Ridge National Laboratory, Oak Ridge, Tennessee 37831, USA}
\affiliation{Physics Division, Oak Ridge National Laboratory, Oak Ridge, Tennessee 37831, USA}

\author{C. Izzo}
\affiliation{Facility for Rare Isotope Beams, East Lansing, Michigan 48859, USA}

\author{A. Jacobs}
\affiliation{Argonne National Laboratory, Lemont, Illinois 60439, USA}

\author{A. Javaji}

\author{S. Kakkar}
\affiliation{TRIUMF, 4004 Wesbrook Mall, Vancouver, British Columbia V6T 2A3, Canada}
\affiliation{Department of Physics and Astronomy, University of Manitoba, Winnipeg, Manitoba R3T 2N2, Canada}

\author{B. Kootte}
\affiliation{TRIUMF, 4004 Wesbrook Mall, Vancouver, British Columbia V6T 2A3, Canada}
\affiliation{Department of Physics and Astronomy, University of Manitoba, Winnipeg, Manitoba R3T 2N2, Canada}

\author{G. Kripko-Koncz}
\affiliation{II.~Physikalisches Institut, Justus-Liebig-Universit{\"a}t Gie{\ss}en, 35392 Gie{\ss}en, Germany}
\affiliation{Helmholtz Research Academy Hesse for FAIR (HFHF), GSI Helmholtz Center for Heavy Ion Research, Campus Gie{\ss}en, 35392 Gie{\ss}en, Germany}

\author{Ali Mollaebrahimi}
\affiliation{GSI Helmholtzzentrum f{\"u}r Schwerionenforschung GmbH, Planckstra{\ss}e 1, 64291 Darmstadt, Germany}

\author{D. Lascar}
\affiliation{Center for Fundamental Physics, Northwestern University, Evanston, Illinois 60208, USA}

\author{E. M. Lykiardopoulou}
\affiliation{TRIUMF, 4004 Wesbrook Mall, Vancouver, British Columbia V6T 2A3, Canada}
\affiliation{Department of Physics and Astronomy, University of British Columbia, Vancouver, British Columbia V6T 1Z1, Canada}

\author{I. Mukul}
\affiliation{TRIUMF, 4004 Wesbrook Mall, Vancouver, British Columbia V6T 2A3, Canada}

\author{S. F. Paul}
\affiliation{TRIUMF, 4004 Wesbrook Mall, Vancouver, British Columbia V6T 2A3, Canada}

\author{W. R. Pla{\ss}}
\affiliation{II.~Physikalisches Institut, Justus-Liebig-Universit{\"a}t Gie{\ss}en, 35392 Gie{\ss}en, Germany}
\affiliation{GSI Helmholtzzentrum f{\"u}r Schwerionenforschung GmbH, Planckstra{\ss}e 1, 64291 Darmstadt, Germany}

\author{W. S. Porter}
\affiliation{TRIUMF, 4004 Wesbrook Mall, Vancouver, British Columbia V6T 2A3, Canada}
\affiliation{Department of Physics and Astronomy, University of British Columbia, Vancouver, British Columbia V6T 1Z1, Canada}
\altaffiliation{University of Notre Dame, Notre Dame, Indiana, 46556, USA}

\author{M. P. Reiter}
\affiliation{School of Physics and Astronomy, University of Edinburgh, Edinburgh EH9 3FD, United Kingdom}

\author{J. Ringuette}
\affiliation{TRIUMF, 4004 Wesbrook Mall, Vancouver, British Columbia V6T 2A3, Canada}
\affiliation{Department of Physics, Colorado School of Mines, Golden, Colorado 80401, USA}

\author{H. Schatz}
\affiliation{Facility for Rare Isotope Beams, East Lansing, Michigan 48859, USA}
\affiliation{Department of Physics and Astronomy, Michigan State University, East Lansing, Michigan 48824, USA}
\affiliation{Joint Institute for Nuclear Astrophysics Center for the Evolution of the Elements (JINA-CEE), East Lansing, Michigan 48854, USA}

\author{C. Scheidenberger}
\affiliation{II.~Physikalisches Institut, Justus-Liebig-Universit{\"a}t Gie{\ss}en, 35392 Gie{\ss}en, Germany}
\affiliation{GSI Helmholtzzentrum f{\"u}r Schwerionenforschung GmbH, Planckstra{\ss}e 1, 64291 Darmstadt, Germany}
\affiliation{Helmholtz Forschungsakademie Hessen f\"ur FAIR (HFHF), GSI Helmholtzzentrum f\"ur Schwerionenforschung, Campus Gie{\ss}en, 35392 Gie{\ss}en, Germany}

\author{C. Walls}
\affiliation{TRIUMF, 4004 Wesbrook Mall, Vancouver, British Columbia V6T 2A3, Canada}

\author{Y. Wang}
\affiliation{TRIUMF, 4004 Wesbrook Mall, Vancouver, British Columbia V6T 2A3, Canada}
\affiliation{Department of Physics and Astronomy, University of British Columbia, Vancouver, British Columbia V6T 1Z1, Canada}

\author{A. A. Kwiatkowski}
\affiliation{TRIUMF, 4004 Wesbrook Mall, Vancouver, British Columbia V6T 2A3, Canada}
\affiliation{Department of Physics and Astronomy, University of Victoria, Victoria, British Columbia V8P 5C2, Canada}

\begin{abstract}
We report precision mass measurements of \isotope{Sr}{74-76} performed with the TITAN multiple-reflection time-of-flight mass spectrometer.
This marks a first time mass measurement of \isotope{Sr}{74} and gives increased mass precision to both \isotope{Sr}{75} and \isotope{Sr}{76} which were previously measured using storage ring and Penning trap methods, respectively.
This completes the $\rm A=74$, $\rm T=1$ isospin triplet and gives increased precision to the $\rm A=75$, $\rm T=1/2$ isospin doublet which are both the heaviest experimentally evaluated triplets and doublets to date.
The new data allow us to evaluate coefficients of the isobaric multiplet mass equation for the first time at $\rm A=74$, and with increased precision at $\rm A=75$.
With increased precision of \isotope{Sr}{75}, we confirm the recent measurement reported by CSRe which was used to remove a staggering anomaly in the doublets.
New \textit{ab initio} valence-space in-medium similarity renormalization group calculations of the $T=1$ triplet are presented at $A=74$.
We also investigate the impact of the new mass data on the reaction flow of the rapid proton capture process in type I x-ray bursts using a single-zone model.
\end{abstract}
\maketitle


Isospin symmetry is based on the observation that protons (p) and neutrons (n) have the same spin, nearly identical masses \cite{tiesinga_2021codata}, and exhibit very similar pp, nn and pn scattering lengths \cite{lam2013isospin}.
Within the isospin framework, protons and neutrons are described as particles distinguished by their isospin projection $\rm T_z$, analogous to projections of angular momentum.
Based on this symmetry, one would assume that isobaric nuclei which share the same nucleon number $\rm A=N+Z$ but differ in their neutron (N) and proton (Z) numbers, exhibit similar structure.
This is indeed true, as the study of nuclear isobars mirrored about the $\rm N=Z$ line is an important avenue for understanding isospin symmetry in nuclei \cite{Bentley_2007coulomb}.
Isospin is however broken by the difference in electric charge and mass between protons and neutrons, and a small contribution from isospin-violating interactions of nuclear origin \cite{ormand_1989empirical,nolen1969coulomb}.
Understanding exactly how these isospin-violating interactions manifest in the structure of nuclei near $\rm N=Z$ has relevance to unitarity tests of the Cabibbo-Kobayashi-Maskawa (CKM) quark-mixing matrix \cite{Hardy_2020superallowed,towner_2008improved}, and the rapid proton capture process (rp-process) in type I x-ray bursts \cite{Ong_2017low,Saxena_202257zn,klochko_2021isobaric}.

A method for studying the nature of isospin symmetry breaking near $\rm N=Z$ is the Isobaric Multiplet Mass Equation (IMME) \cite{maccormick_2014evaluated}.
An isobaric multiplet is a set of nuclear states with the same isospin $\rm T$, and spin-parity $\rm J^{\pi}$, but with differing isospin projections $\rm T_z = (N-Z)/2$.
As long as isospin-violating interactions can be expressed as two-body forces of a tensor of rank two, the IMME states that the mass difference between these isobaric analog states follows a quadratic form:
\begin{equation} \label{eq:IMME}
    \rm ME(\alpha, T, T_z) = a(\alpha,T) + b(\alpha, T)T_z + c(\alpha,T)T_z^2
\end{equation}
where $\mathrm{ME}$ is the mass excess and the other quantum numbers of the system are encapsulated in $\rm \alpha$.
The coefficients $\rm a$, $\rm b$, and $\rm c$ are empirically determined through fitting isobaric analog state data deduced from measurements of nuclear masses and level energies \cite{maccormick_2014evaluated}.
The most recent extensive evaluation of the IMME was performed by MacCormick and Audi who used experimental mass and nuclear level structure data in the $\rm A=10-60$ region to evaluate multiplets with $\rm T=1/2$ up to $\rm 3$ \cite{maccormick_2014evaluated}.



Attempts to theoretically reproduce the $\rm T=1/2$ doublet $\rm b$ coefficients have shown mixed results.
These $\rm b$ coefficients are extracted from ground state mass data of odd-A nuclei and their global trend shows a characteristic staggering pattern between the $\rm A=4n+1$ and $\rm A=4n+3$ nuclei (where $\rm n$ is a positive integer).
This staggering pattern has been explained as due to an increase in Coulomb repulsion between spin anti-aligned protons \cite{feenberg1946theory,Jaenecke_1966vector,janecke1967systematics}.
Evaluation of the $\rm A=69-75$ doublets based on mass data from the 2020 Atomic Mass Evaluation (AME2020) \cite{AME2020_pt1} indicated a confounding departure from the regular staggering pattern.
This staggering anomaly was first noted by Kaneko \textit{et al.} \cite{kaneko_2013variation} who were unable to reproduce it using a modern shell-model approach with inclusion of Coulomb and isospin non-conserving nuclear forces.
Theoretical approaches such as density functional theory (DFT) \cite{bkaczyk_2019isobaric} and valence-space in-medium similarity renormalization group (VS-IMSRG) \cite{li_2023investigation} were also unsuccessful in reproducing this anomaly.
Further discussion of the anomaly is given by Tu \textit{et al.} \cite{tu2014survey}.
It was recently shown by Li \textit{et al.} \cite{Li_2024_Exploring} that the increased mass precision of \isotope{Kr}{71} and \isotope{Sr}{75} reported by Wang \textit{et al.} \cite{wang_2023mass} remove this staggering anomaly.
Wang \textit{et al} only reported four counts of \isotope{Sr}{75} during their experiment, therefore independent confirmation of the mass would strengthen the result of the rectified staggering phase.



Isospin symmetry breaking in nuclei plays an important role in testing CKM matrix unitarity via precise determination of the up-down element $\rm V_{ud}$ \cite{towner_2008improved}.
Extractions of $\rm V_{ud}$ rely on global comparisons of superallowed $\rm \beta$-decay $\rm \mathcal{F}t$-values.
These $\rm \mathcal{F}t$-values are determined from experimentally deduced $\rm ft$-values which are corrected to be nucleus-independent using theoretical corrections for radiative and isospin symmetry breaking effects.
Recent re-evaluations of the radiative corrections have led to an increased deviation in CKM matrix unitarity \cite{seng_2018reduced,seng_2019dispersive,czarnecki_2019radiative}, thus leaving the isospin symmetry breaking corrections as the leading contributor to uncertainty in $\rm V_{ud}$ \cite{Hardy_2020superallowed}. Precise extraction of the isospin symmetry breaking corrections is therefore a major task for this field to progress.

While isospin symmetry breaking corrections for superallowed $\rm \beta$-decays have so far been primarily extracted via shell-model calculations \cite{towner_2008improved}, recent progress in \textit{ab initio} theory has prompted an interest in using alternative methods such as VS-IMSRG theory \cite{martin_2021testing,stroberg_2023talk} and the no-core shell-model \cite{gennari_2023talk}.
\textit{Ab initio} methods hold the key promise of providing a more robust quantification of isospin symmetry breaking corrections, however, careful benchmarking against experimental data is required to make sure the predicted symmetry breaking correction factors meet accuracy demands for CKM unitarity tests.

Comparisons between theoretical and experimental IMME coefficients provide an ideal test bed to probe the reliability of theoretical nuclear forces and the isospin symmetry breaking corrections determined from them.
Recent VS-IMSRG calculations of the triplet $\rm c$ coefficients of all $\rm A=4n+2$ nuclei in the $\rm A=10-74$ range agreed with experimental coefficients within $250$\,keV, but lacked the precision to improve upon available isospin symmetry breaking corrections \cite{martin_2021testing}.
To obtain improved correction terms, refinements of theory guided by further benchmarking against experimental data are needed.


This letter reports on precision mass measurements of neutron-deficient \isotope{Sr}{74-76}.
Our measurement of \isotope{Sr}{75} provides independent confirmation of the storage ring measurement by Wang \textit{et al.}\cite{wang_2023mass}, supporting the removal of the staggering anomaly reported by Li \textit{et al.} \cite{Li_2024_Exploring}.
The new \isotope{Sr}{74} mass value allows us to extend the triplet coefficients into otherwise unexplored regions of the nuclear chart.
Multi-shell VS-IMSRG calculations \cite{Herg16PR,Stroberg_2017nudep,Miya20lMS} at $\rm A=74$ were performed using two new $\chi$EFT-derived interactions to explore the chiral nuclear force.
We also investigated the impact of the new strontium masses on the rp-process using a single-zone type-I x-ray burst model \cite{Schatz_2001}.

The experiment took place at TRIUMF's Ion Trap for Atomic and Nuclear science (TITAN) \cite{dilling2003proposed}, an experimental installation for precision mass measurements and in-trap nuclear decay spectroscopy.
The measurements were performed using TITAN's multiple-reflection time-of-flight mass spectrometer (MR-ToF-MS) \cite{Jesch_2015,dickel2019recent} which provided the means to overcome the low production cross sections and high in-beam contamination typical for this region of the nuclear chart.
The primary radioactive ion beam (RIB) was produced using the isotope separation on-line method \cite{Blumenfeld_2013} at TRIUMF's Isotope Separator and Accelerator (ISAC) facility \cite{dilling2014isac}.
A $50\,\mu$A, $480$\,MeV p$^+$ beam was impinged onto a niobium target to produce the neutron-deficient strontium isotopes of interest.
To enhance the signal-to-background ratio, the strontium isotopes were selectively ionized by TRIUMF's Resonant Ionization Laser Ion Source (TRILIS) using a two-step excitation scheme into an autoionizing state \cite{baig1998,Lassen_forthcoming}.
Following production, ISAC's magnetic mass separator system \cite{dilling2014isac} filtered the RIB down to a range of a single mass unit using a mass separation power of $\rm \frac{m}{\Delta m} \approx 2000$.
The continuous RIB was guided at $20$\,keV into TITAN's helium-filled Radio-Frequency Quadrupole (RFQ) trap \cite{Brunner_2012_NIMA_RFQ} for beam cooling and bunching.
The prepared RIB bunches were lowered to a transport energy of $1.3$\,keV using a pulsed drift tube and supplied at a repetition rate of $50$\,Hz to the buffer-gas-filled RFQ ion transport section of the MR-ToF-MS \cite{Reiter_2021}.

\begin{figure}[h]
    \centering
    \includegraphics[width=0.5\textwidth]{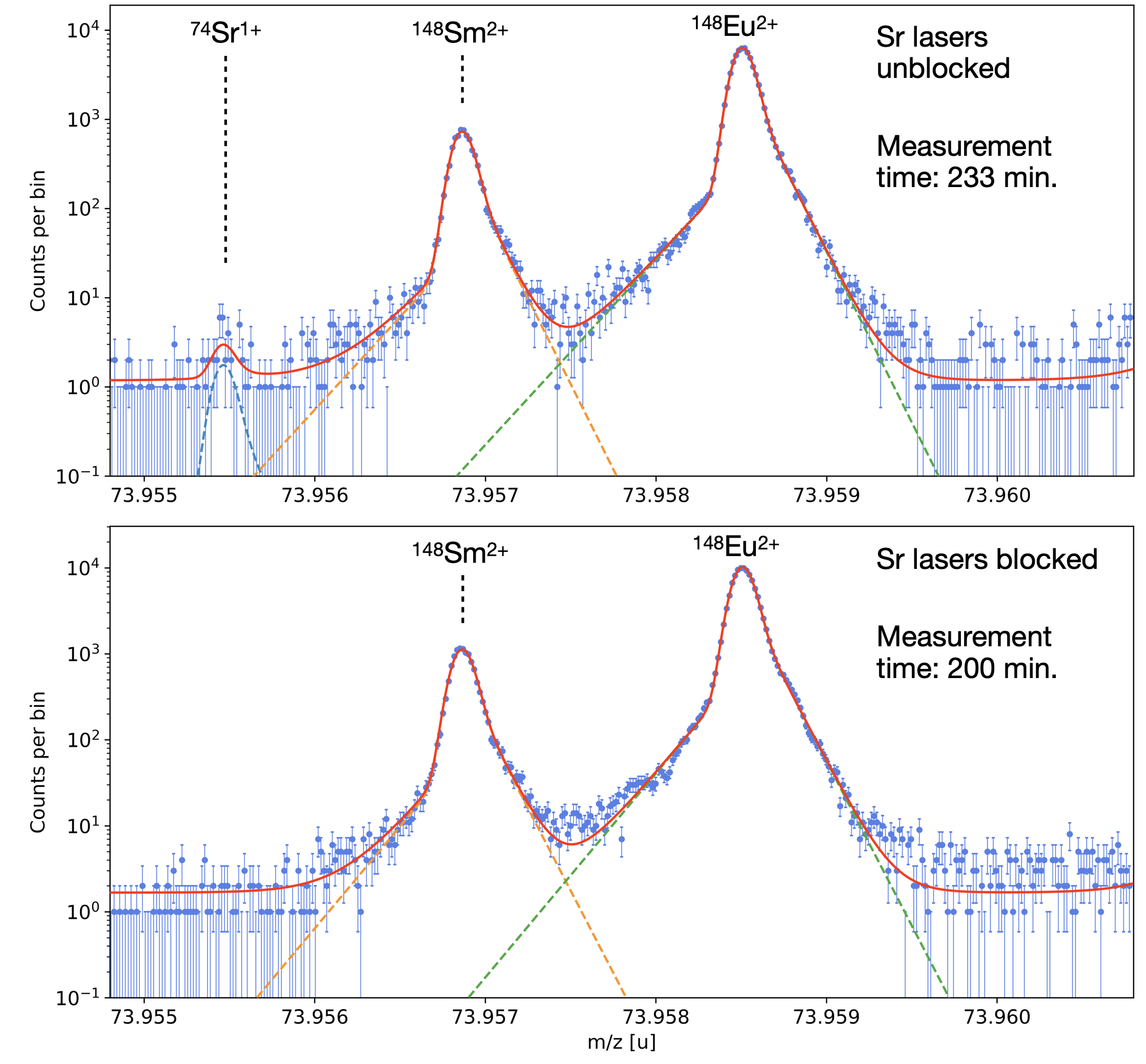}
    \caption{Mass spectra obtained at $A=74$. The spectra have been drift corrected using the time resolved calibration procedure and fitted using Hyper-EMG peak shapes. Error bars are $\rm \pm 1\sigma$ assuming Poisson statistics in each bin. Multi-peak fits to the data (red line) and the underlying single peak fits (dashed lines) are indicated. Unambiguous identification of \isotope{Sr}{74} is demonstrated with the TRILIS lasers which were tuned to selectively ionize only strontium.}
    \label{fig:Sr74_lasers}
\end{figure}

The TITAN MR-ToF-MS consists of a buffer-gas-filled RFQ ion transport section followed by a linear Paul trap and the ToF mass analyzer.
The linear Paul trap prepares and injects cold ion bunches into the mass analyzer, a symmetric assembly of two opposing ion mirrors.
At the end of the ToF analysis, the ions are released through the back mirror onto a MagneTOF\textsuperscript{\textregistered} detector \cite{ETP_magneTOF} which records the ion ToF.
During this experiment, the MR-ToF-MS was operated in a mass-selective retrapping mode \cite{Dickel_2015,dickel2017dynamical}, allowing the device to operate as an isobaric mass purifier and mass spectrometer.
To this end, the ions were injected at two separate stages into the mass analyzer.
The mass-selective re-trapping was done after $20$--$40$ isochronous turns in the mass analyzer.
The ions were then injected back into the mass analyzer for the mass measurement stage which involved $756$--$842$ isochronous laps, resulting in mass resolving powers of $\rm R = \frac{m}{\Delta m} = 400,000-530,000$.
The mass-selective retrapping mode strongly suppressed isobaric contamination, allowing a factor of 10 higher beam intensity during the measurement of \isotope{Sr}{74}.
At each mass unit, the resonant laser ionization scheme was temporarily blocked, resulting in a substantial reduction in Sr yield (see Fig. \ref{fig:Sr74_lasers}), thus verifying the correct identification of isotope in the mass spectrum.
More details of the TITAN MR-ToF-MS can be found in \cite{Reiter_2021}.

During the experiment, each ToF was converted to a mass-to-charge ratio ($\frac{m}{z}$) using the relation \cite{AyetSanAndres_2019},
\begin{equation}
    \rm \frac{m}{z} = \frac{c (t-t_0)^2}{(1 + bN_{it})^2}
\end{equation}
where $\rm t_0$, $\rm b$, and $\rm c$ are calibration parameters, and $\rm N_{it}$ is the number of isochronous turns completed by the species.
The calibration parameters $\rm c$ and $\rm t_0$ were determined before the experiment with reference ions of well-known mass.

After the experiment, a time-resolved calibration was performed to correct for low frequency (sub-Hz) fluctuations in the mirror electrode voltages and thermal expansion of the mass analyzer which cause a fluctuation in the ToF during the measurement.
Time-resolved calibrations were performed for each spectrum according to the procedure outlined in \cite{AyetSanAndres_2019}.
To deduce peak positions and their respective mass values, each spectrum was fitted using the maximum likelihood approach implemented in the emgfit Python package \cite{paul_emgfit}.
The fits were performed using Hyper-Exponentially Modified Gaussian (Hyper-EMG) peak shapes \cite{purushothaman_2017hyper} whose tail order was determined using the procedure described in \cite{Paul_2021}.
After fitting, a final mass calibration was performed using mass peaks in the spectrum with well-known literature values.
Isobaric rubidium ions were used to calibrate \isotope{Sr}{75,76}.
Due to a lack of suitable isobaric reference species, the \isotope{Sr}{74} mass was determined using \isotope{Sm$^{2+}$}{148} as a mass calibrant.
Statistical and systematic mass uncertainties were evaluated following the procedures and methods described in \cite{AyetSanAndres_2019,Paul_2021,paul_emgfit}.
For \isotope{Sr}{75,76}, a non-ideal extraction of the ions from the mass analyzer dominated the mass uncertainty, while for \isotope{Sr}{74} the dominant contribution was statistical uncertainty.



\begin{table*}
\centering
\begin{threeparttable}
    \begin{ruledtabular}
    \begin{tabular}{ccccp{2.5cm}cc}
        Nuclide & $\rm T_{1/2}$ (ms) & $\rm N_{counts}$ & Mass calibrant & \multicolumn{3}{c}{Mass excess (keV)} \\ \cline{5-7}
                &           &              &                & Literature values &      TITAN       & TITAN - Literature \\
        \hline
        $^{74}$Sr &  27.6(2.7) &   36(12)   & $^{148}$Sm$^{2+}$ & -40\,830\#(100\#)\tnote{1}                             & -40\,970(31) & -140(105) \\
        $^{75}$Sr &  88(3)     &   424(22)  &   $^{75}$Rb$^+$   & -46\,620(220)\tnote{1} \newline -46\,200(150)\tnote{2} & -46\,273(10) & 347(220) \\
        $^{76}$Sr & 7\,890(70) & 2\,108(46) &   $^{76}$Rb$^+$   & -54\,250(30)\tnote{1}                                  & -54\,257(9)  & -7(31) \\
    \end{tabular}
    \caption{Summary of mass measurements performed in this work. All masses were measured as singly charged ions. The third column of the table gives $\rm N_{counts}$, which is the number of counts determined by the area of the fitted mass peak.}
    \label{tab:mass_excess_values}
    \end{ruledtabular}
    \begin{tablenotes}
        \item[1] Values from the Atomic Mass Evaluation (AME) 2020 \cite{AME2020_pt2} (\# indicates extrapolations on the mass surface).
        \item[2] Value determined at the CSRe \cite{wang_2023mass}.
    \end{tablenotes}
\end{threeparttable}
\end{table*}

The mass results are summarized in Table \ref{tab:mass_excess_values} with a comparison to literature values.
Our measurement of \isotope{Sr}{76} is in excellent agreement with a Penning trap mass measurement at ISOLTRAP \cite{sikler_2005}, providing independent confirmation of the value and improving the mass precision by a factor of three.
Recently, the experimental cooler storage ring (CSRe) in Lanzhou reported the first direct mass measurement of \isotope{Sr}{75} \cite{wang_2023mass}.
Our measurement is a variance-weighted mean determined from three independent measurements and improves upon the CSRe mass precision by a factor of 15.
Our value is in good agreement with the CSRe value within uncertainty.
Our final mass value of \isotope{Sr}{74} is a variance-weighted mean obtained from two independent measurements.
This constitutes the first time that the ground state mass of \isotope{Sr}{74} has been measured.

\begin{figure}
    \centering
    \includegraphics[width=1.0\columnwidth]{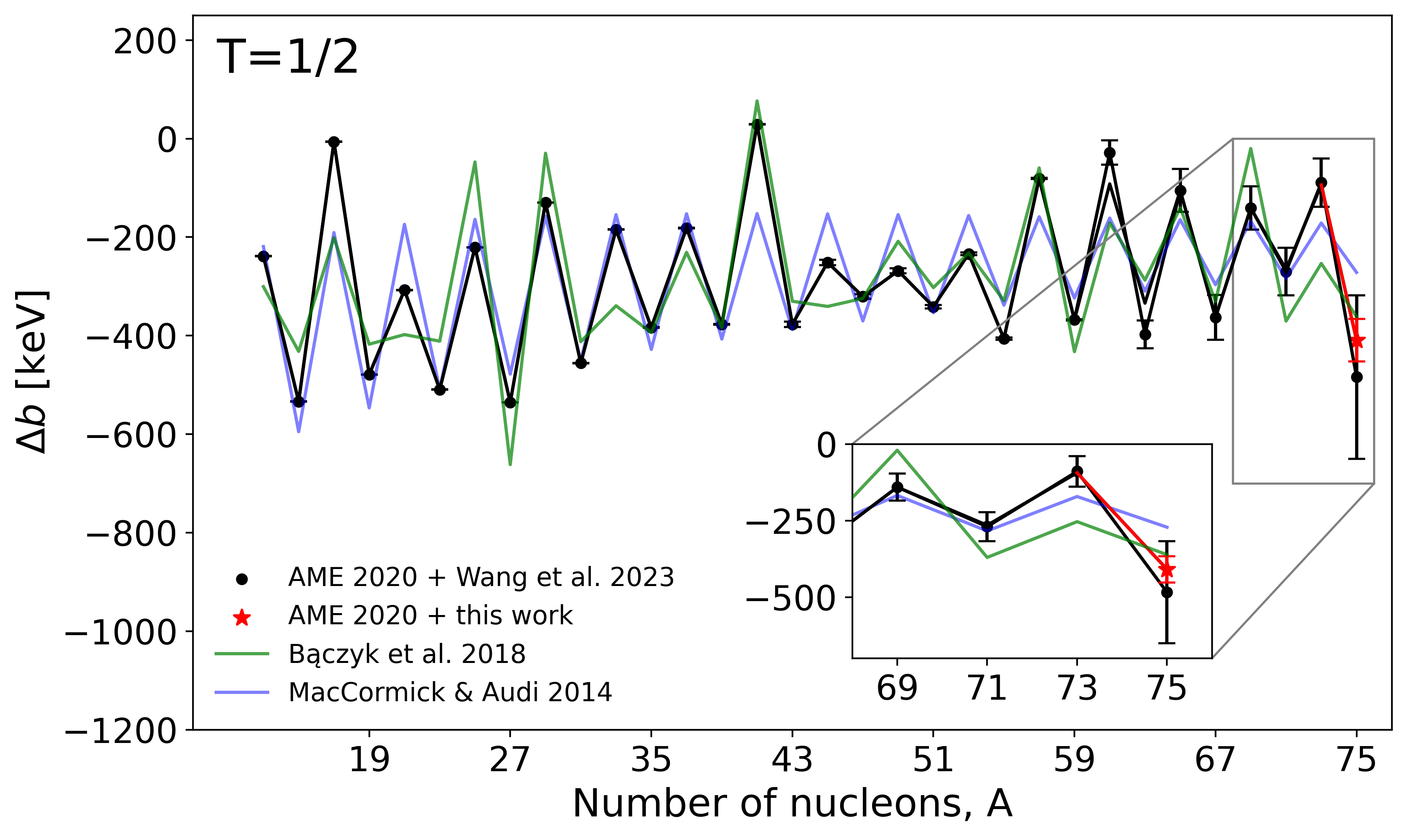}
    \caption{The difference between the doublet $\rm b$ coefficients of adjacent isospin doublets. AME2020 values updated with recent storage ring measurements \cite{wang_2023mass} (black) show the characteristic staggering pattern from $A=10-75$.
    Our measurement of \isotope{Sr}{75} improves upon the precision of the value reported in \cite{wang_2023mass} and confirms the continued staggering phase of $\Delta b$ up to $A=75$.
    For comparison, DFT calculations (green) \cite{bkaczyk_2018isospin} and a modified homogeneously charged-sphere model (blue) \cite{maccormick_2014evaluated} show good agreement.}
    \label{fig:imme_b_coefficient}
\end{figure}

To visualize the staggering behavior of the doublet $b$ coefficients, we consider the quantity,
\begin{equation}
    \rm \Delta b = b(A) - b(A-2)
\end{equation}
which is plotted in Fig. \ref{fig:imme_b_coefficient}.
Data from the AME2020 updated with recent storage ring measurements \cite{wang_2023mass} is shown with the black curve.
As explained in \cite{Li_2024_Exploring}, the mass measurements of \isotope{Kr}{71} and \isotope{Sr}{75} reported in \cite{wang_2023mass} removed the staggering anomaly previously reported in \cite{kaneko_2013variation,tu2014survey}.
Theoretical data from a modified homogeneously charged-sphere approximation \cite{maccormick_2014evaluated} and Density Functional Theory (DFT) \cite{bkaczyk_2019isobaric} do well to reproduce the experimental data.
However, the measurement of \isotope{Sr}{75} reported by \cite{wang_2023mass} was limited to four counts and is therefore dominated by a large uncertainty of $150$\,keV.
As seen from the red curve in Fig. \ref{fig:imme_b_coefficient}, the inclusion of our new \isotope{Sr}{75} value is in good agreement with the previous value and also reduces the uncertainty by a factor of 15.
This in turn supports the rectification of the doublet staggering phase that was shown in \cite{Li_2024_Exploring}.

An interesting feature of the doublet $\rm b$ coefficients is a dampening of the staggering amplitude in the $A=43-55$ $f_{7/2}$ subshell (visible in Fig. \ref{fig:imme_b_coefficient}).
Theoretical models have demonstrated some success in reproducing this feature, although they are in disagreement about the underlying mechanisms responsible for it.
An extended Skyrme $pn$-mixed DFT approach \cite{bkaczyk_2018isospin} reproduced the $f_{7/2}$ dampening, but noted that the inclusion of isospin non-conserving forces increased the staggering amplitude.
A nuclear shell-model approach with modern effective interactions \cite{kaneko_2013variation} also reproduced the $f_{7/2}$ dampening, although the staggering amplitude was instead decreased with the inclusion of isospin non-conserving forces.
In addition to the $f_{7/2}$ dampening, B{\c a}czyk \textit{et al.} \cite{bkaczyk_2018isospin,bkaczyk_2019isobaric} predicted a dampening in the $g_{9/2}$ subshell at $A=83$.
Precision mass measurements at higher masses are needed to test these theoretical indications.


\begin{figure}
    \centering
    \includegraphics[width=1.0\columnwidth]{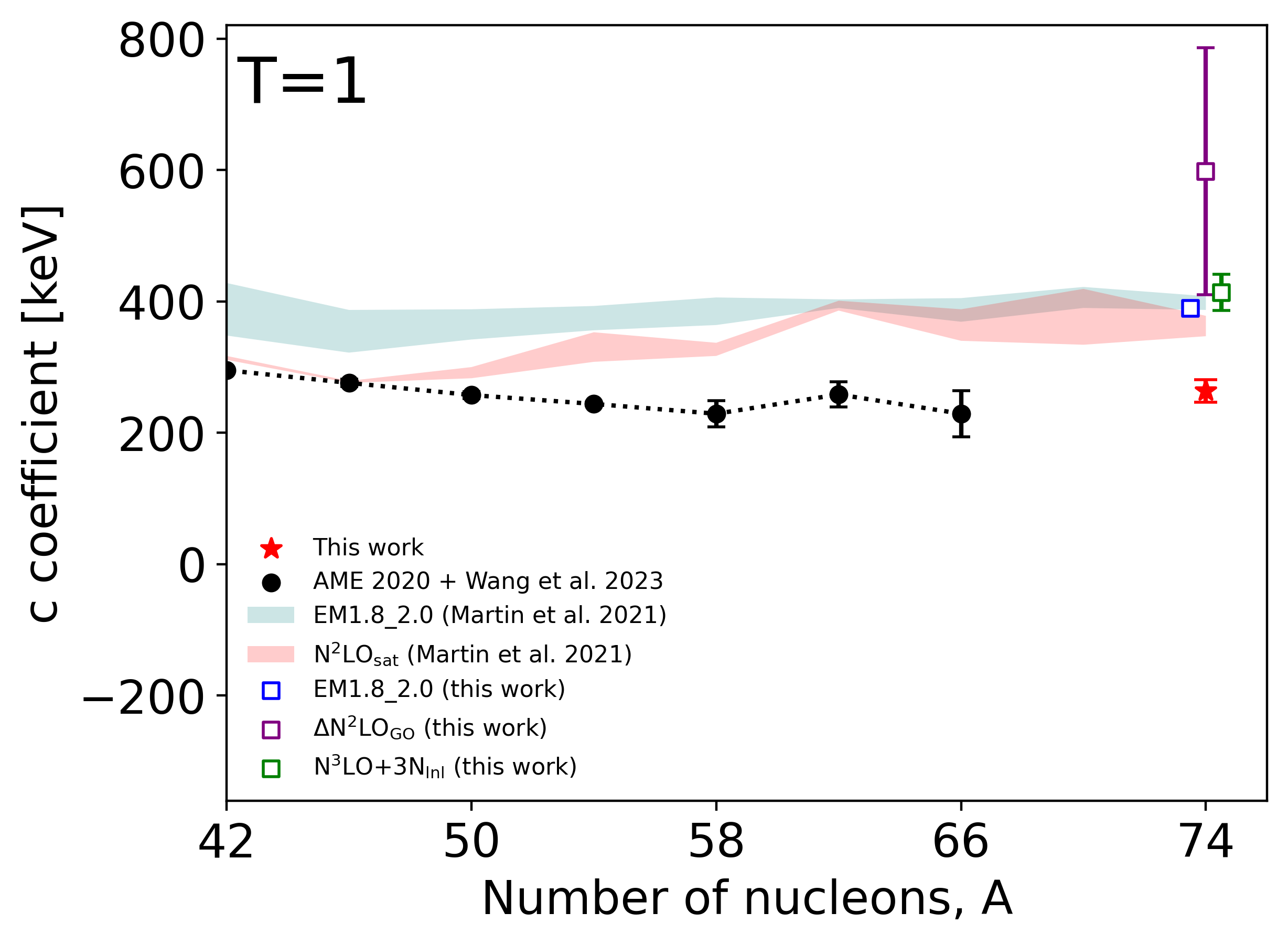}
    \caption{The triplet $\rm c$ coefficients for $\rm A=4n+2$ nuclei. AME2020 values updated with storage ring measurements of \isotope{Zr}{58}, \isotope{Ge}{62} and \isotope{Se}{66} \cite{wang_2023mass} (black) show the relatively constant trend of the $c$ coefficient. Our new value of \isotope{Sr}{74} (red star) shows that this relatively constant trend continues.
    VS-IMSRG calculations (filled curves) from \cite{martin_2021testing} are in good agreement, but consistently over-calculate the coefficient by $50-150$\,keV.
    VS-IMSRG calculations using two new $\chi$EFT-derived interactions at $\rm A=74$ (unfilled squares) do not show a marked improvement.}
    \label{fig:c_coeff_IMSRG}
\end{figure}

Experimentally evaluated triplet data has recently become available up to $\rm A=66$ through direct mass measurements of \isotope{Zn}{58}\cite{wang_2023mass}, \isotope{Ga}{60} \cite{Paul_2021}, \isotope{Ge}{62}\cite{wang_2023mass}, and \isotope{Se}{66} \cite{wang_2023mass,zhou2023mass}.
Experimental evaluations at higher masses have so far been impossible due to a lack of data.
Our measurement of \isotope{Sr}{74} completes the $\rm A=74$ triplet which is now the heaviest experimentally evaluated triplet by eight mass units.
The experimental trend of the triplet $\rm c$ coefficients is shown for $\rm A=4n+2$ nuclei in Fig. \ref{fig:c_coeff_IMSRG} including the new data point.
This confirms that the relatively constant trend displayed by the $\rm A=4n+2$ $\rm c$ coefficients continues to higher masses.


Results from a theoretical VS-IMSRG calculation by Martin \textit{et al.} \cite{martin_2021testing} predict a similarly constant trend, but consistently over-calculate by $50-150$\,keV and diverge near $\rm A=54,58$.
In a follow-up to the calculations by Martin \textit{et al.} \cite{martin_2021testing}, we perform new VS-IMSRG calculations for the $\rm A=74$ triplet.
In the original work, two sets of $\chi$EFT-derived NN+3N interactions were used: $1.8/2.0$(EM)~\cite{Simo17SatFinNuc}, which reproduce ground-state energies across the medium-mass region~\cite{Morr18Tin,Stro21Drip} and N$^2{\rm LO}_{\rm sat}$~\cite{Ekstrom2015}.
The authors used a storage truncation on 3N forces of $\rm e_{\rm 3max} = 16$, but we have extended to $\rm e_{\rm 3max} = 24$ using a novel storage scheme \cite{Miya22Heavy}.
We then explored the chiral nuclear force with two new $\chi$EFT-derived interactions $\Delta$N$^2$LO$_{\rm GO}$~\cite{Jian20N2LOgo} and N$^3$LO+3N$_{\rm lnl}$\cite{Leis18LNL,Soma20LNL}.
Following the same procedure as Martin \textit{et al.} we estimated the IMSRG truncation error using different choices for the reference nucleus: 1) self referenced, 2) $\rm T_z=+1$, 3) $\rm T_z=0$, and 4) $\rm T_z=-1$.

It can be seen in Fig. \ref{fig:c_coeff_IMSRG} that increasing the model space for $\rm 3N$ forces in the $1.8/2.0$(EM) interaction does not significantly improve agreement with the experimental data.
It can also be seen that neither of the two new interactions, $\Delta$N$^2$LO$_{\rm GO}$ and N$^3$LO+3N$_{\rm lnl}$, show a marked improvement.
However, there are still avenues that VS-IMSRG can take to improve agreement with experimental results.
IMSRG(3) is a relatively new iteration of the VS-IMSRG method \cite{Hein21IMSRG3}, which explicitly keeps three-body operators during the flow.
While it is not yet computationally feasible to perform IMSRG(3) at such high masses, it is expected that better agreement with experiment would be demonstrated.

The new strontium mass values also affect the nuclear reaction flows of the rp-process in x-ray bursts.
In particular, the new more precise $^{75}$Sr mass removes mass uncertainties from determining the $^{74}$Rb(p,$\gamma$) vs $\beta$-decay branching.
This branching affects the rp-process production of the p-nucleus $^{74}$Se \cite{Schatz_1998,Weinberg_2006,Herrera2023}, and is thus important for assessing the possibility of the rp-process contributing to the origin of the light p-nuclei.
The improved $^{76}$Sr mass provides an important constraint for the Q-value of the $^{76}$Sr(p,$\gamma$)$^{77}$Y reaction and thus on a possible proton capture bypass of the $^{76}$Sr waiting point in the rp-process \cite{Schatz_1998}.
The precise measurement of the $^{76}$Sr mass is an important first step to reduce the uncertainty of this Q-value. 


To investigate the impact of the new strontium mass values on predictions of type-I x-ray bursts we used a self-consistent single-zone model \cite{Schatz_2001}.
The single-zone model was first benchmarked using reaction rate Q-values determined from current mass uncertainties in the AME2020.
The proton-capture Q-values were extremized by varying the masses of the respective initial and final nucleus by $\pm 3\sigma$. 
The Hauser-Feshbach reaction code TALYS 2.0 \cite{koning_2007talys} was then used to calculate the astrophysical reaction rates \isotope{Rb}{73}$(p,\gamma)$\isotope{Sr}{74}, \isotope{Rb}{74}$(p,\gamma)$\isotope{Sr}{75}, \isotope{Rb}{75}$(p,\gamma)$\isotope{Sr}{76}, and \isotope{Sr}{76}$(p,\gamma)$\isotope{Y}{77}.
The competing photodisintegration rates were calculated from detailed balance.
Included in the calculations is an estimate of the unmeasured \isotope{Y}{77} mass which was calculated from the well-known \isotope{Sr}{77} using the Coulomb displacement energy (CDE) method \cite{Schatz_2017}.
To evaluate the impact of the new mass data on x-ray burst observables, the same procedure was then repeated using the new mass values.


The new masses better pin down the rp-process reaction flow between the \isotope{Kr}{72} and \isotope{Sr}{76} waiting points.
Namely, we find a stronger \isotope{Rb}{74} $\beta$-decay branch and a reduction in mass flowing beyond the \isotope{Sr}{76} waiting point.
The uncertainties induced by nuclear masses for the \isotope{Sr}{76} waiting point bypass \isotope{Sr}{76}(p,$\gamma$) have been reduced from 50\% to 20\%, with the remaining uncertainty arising from the unmeasured \isotope{Y}{77} mass that is needed together with the \isotope{Sr}{76} mass to obtain the important \isotope{Sr}{76} p-capture Q-value. The new masses reduce the \isotope{Sr}{76} bypass flow further from 2\% to just 0.7\%, which leads to an increased production of stable $\rm A=74$ ashes. The previous mass induced uncertainty for the $\rm A=74$ ashes was 16\% and is now a negligible $0.7$\%.
A more precise future measurement of the \isotope{Y}{77} mass would be desirable to confirm these results.


In summary, precision mass measurements of neutron-deficient strontium ions were performed with TITAN's MR-ToF-MS, providing a first time measurement of \isotope{Sr}{74} and improved mass precision of both \isotope{Sr}{75} and \isotope{Sr}{76}.
These measurements complete the heaviest experimentally evaluated isospin doublet measured to date and give improved precision to the heaviest isospin triplet, allowing us to study isospin symmetry breaking using the IMME.
Our measurement of \isotope{Sr}{75} is in good agreement with a storage ring measurement \cite{wang_2023mass} and improves upon the precision by a factor of 15.
This value supports the rectification of the doublet staggering phase that was shown in \cite{Li_2024_Exploring}.
The new data also extend the IMME coefficients for isospin triplets to $\rm A=74$ which is far higher than previously reached and confirms the constant trend of the $\rm c$ coefficients.
New VS-IMSRG calculations at $\rm A=74$ were performed to explore the chiral nuclear force with two new interactions, but did not show a marked improvement over previous calculations by Martin \textit{et al.} \cite{martin_2021testing}.
In a single zone x-ray burst model, the new masses altered the reaction flow in the Sr region significantly, though the resulting impact on observables was found to be relatively small.


TITAN is funded by the Natural Sciences and Engineering Research Council (NSERC) of Canada and through TRIUMF by the National Research Council (NRC).
This work was partially supported by the National Science Foundation under Award Nos. PHY-1430152 (JINA Center for the Evolution of the Elements), OISE-1927130 (IReNA), PHY-2209429, the U.S. Department of Energy, Office of Science, under SciDAC-5 (NUCLEI collaboration), the German Federal Ministry for Education and Research (BMBF)
under contracts no.\ 05P19RGFN1 and 05P21RGFN1, the German Research Foundation (DFG) under contract no.\ 422761894, by HGS-HIRe, and by Justus-Liebig-Universit{\"a}t Gie{\ss}en and GSI under the JLU-GSI
strategic Helmholtz partnership agreement. 
The IMSRG code used is ragnar$\_$imsrg \cite{ragnar}.
Theoretical efforts were further supported by NSERC under grants SAPIN-2018-00027, RGPAS-2018-522453 and the Arthur B. McDonald Canadian Astroparticle Physics Research Institute. Computations were based on the performed with an allocation of computing resources on Cedar at WestGrid and Compute Canada.
For the purpose of open access, the author has applied a Creative Commons Attribution (CC BY) licence to any Author Accepted Manuscript version arising from this submission.



\newcommand{\noop}[1]{}

\end{document}